\documentclass[runningheads]{llncs}
\usepackage[T1]{fontenc}
\usepackage{graphicx}
\usepackage{todonotes}
\usepackage{enumitem}
\usepackage{microtype}
\usepackage{orcidlink}
\usepackage[table]{xcolor}
\usepackage{multicol} % needed for the list of participants, index
\usepackage{aliascnt}
\usepackage{booktabs}
\usepackage{array}

\begin{document}
\title{\textit{ContinuumConductor}: Decentralized Process Mining on the Edge-Cloud Continuum}
%\title{Embedded Process Mining: Gaining operational process insights on decentralized and unstructured data streams}
%
\titlerunning{Decentralized Process Mining on the Edge-Cloud Continuum}
\author{
Hendrik Reiter\inst{1} \orcidlink{0009-0003-8544-0012} \and
Janick Edinger\inst{2} \orcidlink{0000-0002-9392-2922} \and 
Martin Kabierski\inst{3} \orcidlink{0000-0002-9852-7489} \and
Agnes Koschmider\inst{4} \orcidlink{0000-0001-8206-7636} \and
Olaf Landsiedel\inst{1,}\inst{5} \orcidlink{0000-0001-6432-300X} \and
Arvid Lepsien\inst{1} \orcidlink{0000-0002-8105-382X} \and
Xixi Lu\inst{6}\orcidlink{0000-0002-9844-3330} \and
Andrea Marrella\inst{7} \orcidlink{0000-0002-1031-0374} \and
Estefania Serral\inst{8} \orcidlink{0000-0001-7579-910X} \and
Stefan Schulte\inst{5} \orcidlink{0000-0001-6828-9945} \and
Florian Tschorsch\inst{9} \orcidlink{0000-0001-6716-7225)} \and
Matthias Weidlich\inst{10} \orcidlink{0000-0003-3325-7227} \and
Wilhelm Hasselbring\inst{1} \orcidlink{0000-0001-6625-4335}
}
%4, https://www.cau-se.de/

\authorrunning{H. Reiter et al.}

\institute{
Kiel University, Kiel, Germany
\\\email{\{hendrik.reiter,hasselbring\}@email.uni-kiel.de}
\and 
University of Hamburg, Hamburg, Germany
\and
University of Vienna, Vienna, Austria
\and
University of Bayreuth, Bayreuth, Germany\\
\and 
Hamburg University of Technology, Hamburg, Germany
\and 
Utrecht University, Utrecht, Netherlands
\and 
Sapienza University of Rome, Rome, Italy
\and 
KU Leuven, Leuven, Belgium
\and 
Dresden University of Technology, Dresden, Germany 
\and 
Humboldt University of Berlin, Berlin, Germany
%\email{marrella@diag.uniroma1.it}
%\email{olaf.landsiedel@tuhh.de}
%\email{x.lu@uu.nl}
%\email{estefania.serralasensio@kuleuven.be}
%\email{florian.tschorschl@tu-dresden.de}
%\email{agnes.koschmider@uni-bayreuth.de}
%\email{martin.kabierski@univie.ac.ate}
%\email{ale@cs.uni-kiel.de}
%\email{matthias.weidlich@hu-berlin.de}
%\email{janick.edinger@uni-hamburg.de}
}

\maketitle              % typeset the header of the contribution
\begin{abstract}
Process mining traditionally assumes centralized event data collection and analysis. However, modern Industrial Internet of Things (IIoT) systems increasingly operate over distributed, resource-constrained edge-cloud infrastructures. This paper proposes a structured approach for decentralizing process mining by enabling event data to be mined directly within the IoT system's edge-cloud continuum. We introduce \textit{ContinuumConductor} a layered decision framework that guides when to perform process mining tasks such as preprocessing, correlation, and discovery centrally or decentrally. Thus, enabling  privacy-preserving, responsive and resource-efficient process mining. For each step in the process mining pipeline, we analyze the trade-offs of decentralization versus centralization across these layers and propose decision criteria. We demonstrate \textit{ContinuumConductor} at a real-world use-case of process optimazition in inland ports.
Our contributions lay the foundation for computing-aware process mining in cyber-physical and IIoT systems.

\end{abstract}

\section{Introduction}

The proliferation of sensors and actuators forms the backbone of modern Industrial Internet of Things (IIoT) environments. These systems leverage vast amounts of sensor data to monitor processes, while actuators perform actions often in direct response to the insights derived from this data. This complex interaction requires the design of robust and efficient computing architectures that can meet three critical objectives: firstly, responsive analysis ensuring that data analysis is performed in real-time to allow timely actions; secondly, privacy-preservation, particularly important in scenarios involving human interaction where sensitive behavioral data, such as that captured by cameras, must be handled with care and third resource-efficiency, since transferring and processing large data volume may exceed the devices capacities.

The edge-cloud continuum~\cite{Satyanarayanan2017} offers a promising solution to achieve real-time responsiveness, enhanced privacy protection and resource-efficiency. By distributing computational tasks closer to the data source (edge) while leveraging the scalability of centralized cloud resources, these architectures can mitigate latency, reduce the exposure of sensitive information and minimize transferred data.
In this setting, process mining stands out as a powerful technique. Traditionally applied to centralized event logs for retrospective business insights, process mining in the dynamic, data-rich IIoT context requires a full pipeline, from preprocessing raw sensor data to visualizing processes and extracting actionable insights.
%In this context, process mining emerges as a powerful techniques where traditionally, the algorithms are applied to a previously captured and centralized event log, providing retrospective insights into business processes. However, in the dynamic and data-intensive realm of IIoT, the application of process mining involves a comprehensive pipeline, capturing steps from the initial preprocessing of raw sensor data to the final visualization of processes and extraction of actionable insights.

This paper delves into the benefits and challenges of transforming the conventional process mining pipeline into a distributed paradigm across the edge-cloud continuum. Specifically, this paper contributes by:

\begin{enumerate}
    \item 
    Presenting a real-world use case that highlights the practical relevance and requirements of decentralized process mining in an IIoT setting. 
    \item Discussing the key challenges inherent in implementing decentralized process mining on unstructured IoT sensor data streams.
    \item Introducing \textit{ContinuumConductor} a decision framework for the placement of computational steps within the process mining pipeline, determining where computations should be executed within the edge-cloud-continuum.
\end{enumerate}

The remainder of this paper is structured as follows: Section 2 describes the problem by introducing the use-case of the automation in inland ports, describing its additional goals for process mining at the edge-cloud-continuum as well as related work to achieve them. Section 3 demonstrates the process mining pipeline. Moreover, it proposes techniques how the steps of the process mining pipeline can be executed within the edge-cloud-continuum. Section 4 presents \textit{ContinuumConductor} the decision framework for the placement of those steps and applies it on the use case. Section 5 concludes the paper.  
\section{Problem Description}
We illustrate the need for decentralized process mining by first outlining a specific use-case (\autoref{sec:use_case}), from which then derive general requirements (\autoref{sec:requirements}). Then, we review related work in the light of these requirements (\autoref{sec:related_work}).

\subsection{Use-case: Decentralized Process Mining in Inland Ports}
\label{sec:use_case}

To illustrate the benefits of decentralized process mining, we consider the \emph{InteGreatDrones} project~\cite{Teegen2024}, which aims to modernize data collection and operational transparency in inland port terminals. Inland multi-purpose terminals often operate in highly dynamic environments, characterized by frequent changes in cargo types, varying throughput, and a limited degree of process standardization. As a result, systematic monitoring and process optimization are challenging. 
%particularly in the absence of extensive fixed infrastructure and digitized workflows.

In this use case, a sensor ecosystem is deployed across the terminal to capture fine-grained operational data. The system combines heterogeneous data sources:
\textit{1.~Fixed Cameras:} Permanently installed cameras cover predefined areas of the terminal, capturing video streams that document incoming and outgoing goods as well as vehicle movements. Due to their stable network connections, these cameras can stream high-resolution data directly to central edge servers.
\textit{2.~Vehicle-Mounted Cameras:} Mobile terminal vehicles, such as reach stackers, straddle carriers, and trucks, are equipped with cameras that record their activities and immediate surroundings. These data streams provide context on the movement of cargo and the utilization of terminal equipment but are only intermittently connected to the terminal's IT infrastructure via wireless links.
\textit{3.~Autonomous Drone Cameras:} A fleet of drones autonomously patrols the terminal area, generating aerial video data to monitor operational zones that are otherwise difficult to cover. Drones can dynamically focus on areas of interest, for example, to track specific handling operations or perform targeted inspections.
\textit{4.~Sensor Boxes on Vehicles:} Selected vehicles are equipped with sensor boxes that record GPS position, acceleration, vibration, and the current height above ground of the spreader beam. These readings enable a precise reconstruction of vehicle behavior and cargo handling sequences.
Data processing in this use case has to cope with \emph{heterogeneous connectivity} and \emph{massive data volumes}. While some sensors (e.g., fixed cameras) provide continuous data streams over wired connections, others (e.g., 
drones) rely on variable wireless connectivity. To mitigate bandwidth constraints and latency, sensor nodes perform \emph{local preprocessing} on edge computing resources, which includes data filtering, aggregation, anonymization, and transformation into structured intermediate representations. For example, drone video streams are locally analyzed to extract object trajectories and anonymize sensitive information before transmitting results to the central infrastructure.

The sensor data is consolidated in a middleware platform deployed across the terminal's edge--cloud continuum. From this platform, \emph{event logs} are generated that describe the lifecycle of each cargo unit, including timestamps for arrival, intermediate handling steps, storage movements, and final departure. Moreover, the activities and states of vehicles, such as loading, unloading, idle time, and maintenance-related events, are captured.
Based on these logs, process mining can improve situational awareness, process compliance, and operational efficiency.
%with e.g., a live dashboard on the current status of cargo and resources.
%, predictive analytics for maintenance scheduling, and retrospective process discovery to identify process bottlenecks.
%or deviations from planned workflows. 

\subsection{Goals for Process Mining on the Edge-Cloud-Continuum}
\label{sec:requirements}
From the above application scenario, we derive three additional goals for the process mining analysis:  \\
%\begin{enumerate}[nosep, leftmargin=1.2em]
\textbf{G1) Privacy preservation:} Sensitive data must be anonymized close to the source to comply with privacy regulations and to maintain stakeholder trust.\\
\textbf{G2) Real-time responsiveness:} Immediate detection of process deviations (e.g., unauthorized access) requires near-sensor computation.\\
\textbf{G3) Resource efficiency:} Raw sensor data from high-resolution video and telemetry streams exceed available network bandwidth if transmitted unprocessed.
%\end{enumerate}

These requirements can be fulfilled by employing the edge-cloud continuum and decentralized process mining techniques. By performing process mining tasks closer to the data sources with cloud-based aggregation and analytics, the approach balances latency, data protection, and process insight in a dynamic, resource-constrained environment. Nevertheless, for each step within the pipeline it has to be discussed where the task are placed within the edge-cloud continuum.

\subsection{Related Work}
\label{sec:related_work}

\textbf{Privacy preservation.} 
Privacy considerations in process mining received considerable attention in recent years, especially for IIoT applications~\cite{michael_user-centered_2019}. To address these privacy risks, encryption techniques can facilitate confidentiality~\cite{DBLP:conf/simpda/RafieiWA19} and a multitude of data sanitization techniques have been proposed, adopting group-based privacy 
notions~\cite{DBLP:journals/dke/FahrenkrogPetersenAW23}
%DBLP:journals/dke/RafieiA21} 
or differential privacy~\cite{DBLP:conf/icpm/ElkoumyPD21}.
%~\cite{DBLP:journals/biseMannhardtKBWM19,DBLP:conf/icpm/Fahrenkog-Petersen21,DBLP:conf/icpm/ElkoumyPD21}.
These techniques are not limited to the control-flow of a process, but may be lifted to contextual information contained in an event log~\cite{DBLP:conf/caise/HildebrantFWR23}. 
Most of the existing techniques for protecting privacy in process mining have not been designed for distributed environments that continuously produce event data. 
However, some notable proposals include the use of multiparty computation~\cite{DBLP:conf/icpm/ElkoumyPD21} or distributed cryptography~\cite{DBLP:conf/icws/ZhangKZLC24} for simple process mining tasks and control-flow abstractions of distributed event logs~\cite{DBLP:journals/access/RafieiA23}.
%and sharing of process models among several parties~\cite{DBLP:journals/tsc/LiuDZZLC19}.
%, or the evaluation of privacy under continuous data releases~\cite{DBLP:conf/coopis/RafieiEA22}
%The question of how to comprehensively exploit the edge--cloud continuum to achieve privacy-preserving process mining on continuous data streams is largely open. 

\smallskip
\noindent
\textbf{Real-time responsiveness.} Real-time considerations in process mining are addressed in the field of streaming process mining~\cite{Burattin2022}. Streaming process mining algorithms perform on continuously generated, potentially infinite event streams instead of complete and static event logs. These algorithms require bound runtime and memory usages. Utilizing techniques such as filtering, sampling or windowing process mining techniques process discovery, conformance checking and process enhancement have been transferred to a streaming domain. Although initial approaches~\cite{evermann_scalable_2016,edgeminer} exist to perform streaming process mining tasks in a distributed manner, the approach to perform them within the computing infrastructure close to the data source still lacks further research.

\smallskip
\noindent

\textbf{Resource Efficiency and Distributed Mining.} In process discovery, van der Aalst~\cite{vander_aalst_decomposing_2013} introduced a method for computing and merging partial Petri nets from partial event logs. Techniques like Map-Reduce~\cite{evermann_scalable_2016} parallelize discovery algorithms across multiple compute nodes to manage large event logs. Similarly, for conformance checking, the Single-Entry Single-Exit approach breaks down large models and logs into smaller sub-processes for independent analysis~\cite{MUNOZGAMA2014102}, distributing the workload.
While these methods efficiently handle large event logs, they remain centralized, lacking both scalability and real-time responsiveness. Recently, EdgeMiner~\cite{edgeminer} introduced a distributed, resource-efficient algorithm that operates directly on resource-constrained sensor nodes with real-time event data. This demonstrated the feasibility of near real-time process mining, though it's currently limited to classic footprint-matrix-based algorithms like the alpha miner.

\section{IIoT Process Mining on the Edge-Cloud Continuum}

The use of process mining allows to get insights for efficiently coordinating objects (drones, vehicles) in the port terminal use-case, in order to, for example,
shorten the loading and unloading process or to predict the readiness or deviation of objects before breakdown. However, due to the nature of the distributed IoT, e.g., video data stream used in this use case, several challenges must be addressed to apply process mining efficiently. In this section, we use the use-case presented earlier to discuss these challenges as well as the opportunity the edge cloud continuum brings for enabling goals such as privacy-preservation and real-time analysis. 

\subsection{IIoT Process Mining Pipeline}

\begin{figure}[htbp] 
    \centering
    \setlength{\belowcaptionskip}{-20pt}
    \includegraphics[width=0.8\textwidth]{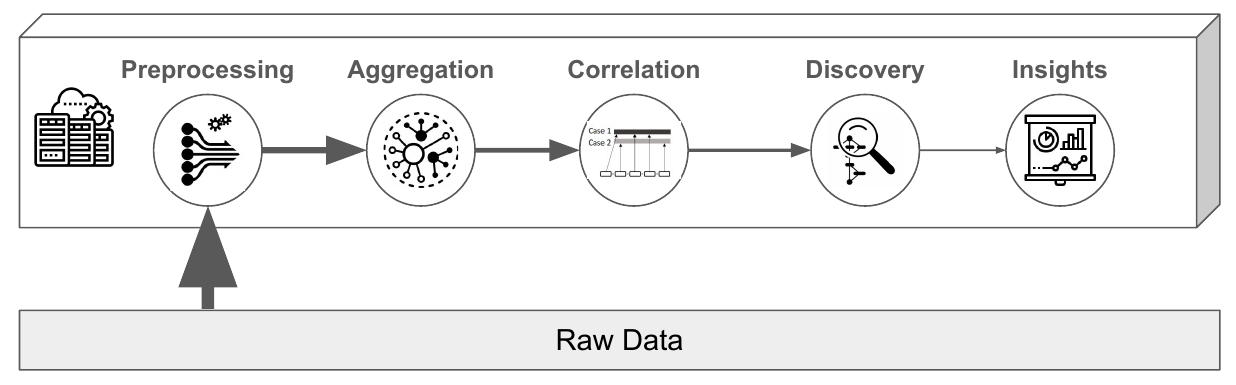}
    \caption{Process Mining Pipeline. Raw sensor data is preprocessed to low-level events. These are aggregated to high-level events and correlated with case/object ids. Further, process models are discovered and insights are extracted from them.}
    \label{fig:pipeline}
\end{figure}

To transform raw unstructured data streams into valuable process insights, the steps of \textit{preprocessing}, \textit{aggregation}, \textit{correlation}, \textit{discovery} and \textit{insights} have to be performed~\cite{DBLP:conf/modellierung/KoschmiderAFILA24}. Figure \ref{fig:pipeline} shows these steps which we refer to as the process mining pipeline in the following.

\textit{1. Preprocessing:}
Usually, IoT data is unstructured and heterogeneous (C1), and exists at a lower level of abstraction compared to the event data traditionally used as input for process mining. IoT devices (i.e., cameras, drones and sensor boxes) generate various data formats such as JSON, video, or time-series sensor readings, which cannot directly be applied by process mining techniques. Although numerous methods exist to abstract unstructured data into a structured form such as an event log, these methods are not generalizable and must be adapted to the specific application purpose~\cite{vanZelst:821230}.
Transforming video data into recognized (low-level) activities computationally complex machine learning, which may exceed compute capabilities (C4), need to be employed~\cite{Lepsien2023}. Due to the probabilistic output of such models an uncertainty (C2) is introduced~\cite{lepsien_ranking_2025}. Moreover, data quality issues~\cite{DBLP:journals/jdiq/HofstedeKMAFSWCWGMS23} such as noise increase this uncertainty further. 
Distribution may be instrumental in reducing uncertainty. For instance, the use of multiple complementary or redundant sensors may be used to resolve ambiguity and mitigate concerns about data quality by covering previously unrecorded parts of the process and ensuring data availability in the presence of sensor failures~\cite{DBLP:conf/modellierung/KoschmiderAFILA24}.\\
\indent\textit{2. Aggregation:}
It is crucial that the aggregation process takes into account the semantics of events in the form of temporal patterns as well as the context (e.g., operating mode of the cameras, shift time of the sensor boxes). Without considering context, the same event can represent different activities (C3). Processing distributed IoT data at the edge in terms of storage, pre-processing, and real-time is also very computationally  heavy (C4). 
Additionally, sensor readings can be inaccurate, incomplete, delayed, or lost, which can lead to incorrect process models or misinterpretations (C5)~\cite{bertrand2025challenges}. \\
\indent\textit{3. Correlation:}
The correlation of recognized activities to specific cases or objects, such as within object-centric process mining~\cite{DBLP:conf/sefm/Aalst19} frameworks, requires a global shared notation (C6), especially in distributed environments. At this stage, the temporal dimension and sequential order of events become paramount (C5). Distinguishing the precedence of events is fundamental for accurate activity correlation. This temporal ordering enables the construction of direct-follow relations within individual cases or objects, a prerequisite for numerous process discovery algorithms. To address the completeness of available data~\cite{DBLP:journals/is/KabierskiRW25}, dynamic windowing techniques can be used to balance the trade-off between accuracy and responsiveness~\cite{Imenkamp2025}.\\
\indent\textit{4. Discovery:} The task of transforming an event log towards a process model is called process discovery. Process discovery algorithms transform underlying process behavior into abstract formal representations such as Petri nets or process trees. The models capture control-flow relations and provide a formal basis for further analysis. In a decentralized setting~\cite{DBLP:conf/smds/Aalst21} multiple actors/organziations can discover process fragments. If the algorithms require a complete event log central processing is beneficial while some algorithms such as the Inductive Miner may work better on smaller, localized event logs (C5).
In terms of privacy, the challenge is to merge local fragments and clarify interfaces (C6) to other actors and organizations to create a whole view of the complete process. \\
\indent\textit{5. Insights:} The final step of the process mining pipeline is extracting insights from the process model. This includes techniques such as visualization, conformance checking, performance diagnostics (service times, waiting times), root-cause analysis, simulation (e.g., digital twin), or operations research (e.g., resource planning)~\cite{vanderAalst2022}. Some techniques require computationally complex machine learning algorithms to be executed on specialized hardware (C4). 
In a distibuted setting each actor may have a partial view of the data due to e.g. access rights. Here, insight extraction must be robust to incomplete data (C5).

\setlength{\tabcolsep}{5pt} 
\begin{table}
    \label{tab:pipeline}
    \centering
    \caption{Challenges and Goals of IIoT process mining}
    \begin{tabular}{l|l}
    \noalign{\hrule height 1.5pt} 
    \textbf{Challenges} & \textbf{Goals} \\
    \noalign{\hrule height 1.5pt} 
    C1) Large volume of unstructured data &  G1) Privacy preservation \\
    C2) Uncertainty &  G2) Real-time responsiveness \\
    C3) Sensitivity to ambiguous context &  G3) Resource efficiency \\
    C4) Network and computing limitations & \\
    C5) Erroneous and incomplete data &  \\
    C6) Necessity of shared case/object notion &  \\
\end{tabular}
\end{table}
\vspace{-10pt}

\subsection{Edge-Cloud Continuum}
\label{sub:continuum}

Edge Computing is a computing paradigm that brings computing and storage closer to data sources of sensors or mobile devices~\cite{Satyanarayanan2017}. For this, the computing capabilities of the end devices, IoT gateways, the network infrastructure, or local micro data centers are utilized. Edge computing promises responsive and efficient services due to lower network latencies and network utilization. Additionally, privacy can be improved by not sharing all data immediately at a central place. %Rather privacy-critical data is stored decentralized within trust domains.  
In contrast, the cloud provides centralized, scalable, and high-performance computing infrastructure capable of handling large-scale data storage. The edge-cloud continuum spans a hierarchical tree of computing resources out of edge, cloud, and fog (in between-edge and cloud) nodes. Each node can either store or compute locally on-device, on a higher computing tier, or on the same computing tier in a peer-to-peer manner.   
The continuum between edge and cloud enables a more efficient, responsive, and privacy-aware process mining architecture. It supports real-time analytics at the edge while leveraging the cloud for deeper, long-term insights, thereby achieving a balance between latency, resource usage, bandwidth consumption, and data privacy \cite{meuser2024}.

In terms of process mining, edge nodes can perform initial processing tasks like data filtering, aggregation, or even lightweight mining.
The cloud can perform more comprehensive analyses, integrate data from multiple edge sources, and apply advanced algorithms for process discovery and conformance checking when compared to edge devices. 

\begin{figure}[htbp] 
    \centering
    \setlength{\belowcaptionskip}{-10pt}
    \includegraphics[width=0.8\textwidth]{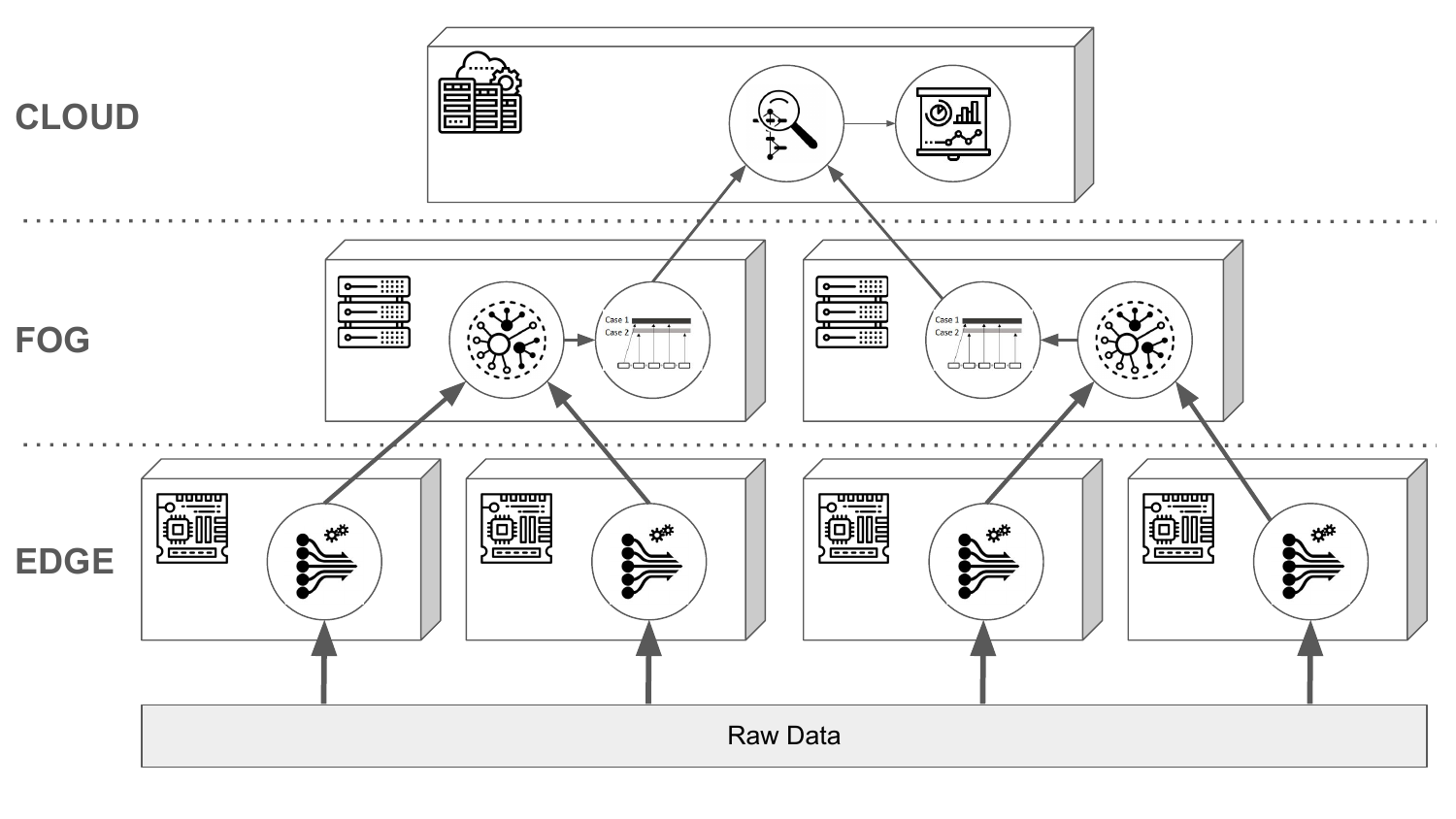}
    \caption{Process Mining Pipeline placed on the edge-cloud continuum. The preprocessing is performed distributedly on the edge nodes. On fog nodes, the data gets abstracted and correlated to cases. This results are then centrallized to perform discovery and insight extraction.}
    \label{fig:deployment}
\end{figure}

\subsection{Goal-Supporting Techniques}

\textbf{Privacy-Preservation.}
In order to determine whether centralized processing is appropriate or if
decentralization is required, privacy threat modeling plays a vital role.
Frameworks such as LINDDUN~\cite{linddun} support this process by
incorporating attacker models and introducing the notion of trust zones,
i.e., areas where data processing is considered safe based on trust
assumptions. Beyond these zones, data sharing may require additional
safeguards or be avoided altogether.

In addition to threat modeling, privacy design strategies offer
guidance for implementing privacy-preserving processing~\cite{hoepman14privacy}.
While applicable to both centralized and decentralized
settings, several technical principles, such as separate, minimize, and
aggregate, naturally align with decentralized architectures. Processing data
closer to its source, as enabled by edge-cloud infrastructures, facilitates
early minimization and aggregation, reducing data exposure and granularity
before transmission. Separation can be achieved by limiting data merging
across trust zones or device boundaries, helping to mitigate linkage risks.
% When applied directly on end devices, these strategies highlight the privacy
% advantages of decentralized processing.
The hide principle further supports broader privacy goals such as
confidentiality and anonymity. Privacy-enhancing technologies~(PETs) like
secure multi-party computation~\cite{yao82mpc} and local differential
privacy~\cite{erlingsson14rappor}, which often rely on distributed
architectures, achieve these goals by processing sensitive data locally and
sharing only obfuscated or aggregated outputs. This further underscores the
privacy benefits of decentralized processing.

\textbf{Real-time and Resource Efficient Analysis.} Edge-side real-time data preprocessing and complex event processing (CEP)~\cite{DBLP:conf/icpm/ImenkampA0K24} are critical to reduce data volume and enable instant insights. By performing filtering, aggregation, and transformation directly at the edge, the amount of data sent to the cloud is drastically reduced. Next, robust, low-latency data ingestion and stream processing frameworks~\cite{evermann_scalable_2016} ensure a swift and reliable event transfer, maintaining real-time fidelity. Incremental process discovery and online conformance checking algorithms are also essential, allowing process models to continuously update and enabling immediate deviation detection. Furthermore, accurate time synchronization across the continuum ensures reliable event ordering, which is fundamental for data integrity. Finally, activity recognition and event abstraction using on-edge AI/ML~\cite{DBLP:journals/ccr/DingPMABDHKLMMO22} improves insight quality. This method generates richer event logs from sensor data, providing precise and actionable real-time insights with the resource efficiency of TinyML.
\section{Continuum Conductor}
% Maybe Task Placement
In previous sections, we gathered challenges, goals and techniques associated with distributing process mining tasks across an Edge-Cloud Continuum. To facilitate decision-making regarding the optimal placement of pipeline steps for individual use cases, we introduce the \textit{ContinuumConductor}. 

\textbf{Framework:} The \textit{ContinuumConductor} is a decision framework designed to determine whether each step in a process mining pipeline should be executed centrally or in a distributed manner. 
Comprising 16 questions that address the previously identified challenges, the current version of the ContinuumConductor aligns with each of the five steps of the process mining pipeline. This tool does not aim to offer a comprehensive list of all potential challenges. Instead, it is designed to initiate a discourse and provide a foundational framework for exploring further dimensions.
The model operates by posing a series of questions, each with four possibilities to perform the compute: centralized (critical), centralized (favorable), decentralized (favorable), decentralized (critical). Table \ref{tab:rtpm_questions} presents the \textit{ContinuumConductor} questions.
A unique scenario arises when both a \textit{`critical centralized'} and a \textit{`critical decentralized'} evaluation are present within the assessment of a single pipeline step. This represents a conflict, necessitating the application of specialized algorithms or hardware adaptations to resolve the architectural dilemma.
To address these conflicts, we present two examples. First, when raw data is privacy-sensitive but on-device processing is too slow, a better solution is to deploy more powerful hardware closer to the device. Second, if a process mining algorithm requires a complete event log, but parts of the model are privacy-critical, it is necessary to design new algorithms. New algorithms require the quantification and optimization of the privacy-utility trade-off based on the specific application's requirements.

\begin{table}[b!]
\centering
\setlength{\abovecaptionskip}{-15pt}
\caption{\textit{ContinuumConductor} questions to decide on the placement of every step in the process mining pipeline within the cloud-edge continuum.}
\label{tab:rtpm_questions}
\begin{tabular}{|>{
\raggedright\arraybackslash}p{2.2cm}| 
>{\raggedright\arraybackslash}p{8cm}|
>{\centering\arraybackslash}p{1cm}|
}
\hline
\textbf{Phase} & \textbf{Question} & \textbf{Chal-lenge} \\
\specialrule{1.5pt}{0pt}{0pt}

\textbf{Preprocessing}   & \textbf{Pre1.} Are compute resources enough for preprocessing? & C1 \\
\arrayrulecolor{black}\cline{2-3} 
 Raw Data $\rightarrow$ & \textbf{Pre2.} Is raw data privacy-critical? & G1\\
\arrayrulecolor{black}\cline{2-3} 
Low-Level & \textbf{Pre3.} Does raw data transfer need high bandwidth? & C4,G3 \\
\arrayrulecolor{black}\cline{2-3} Events & \textbf{Pre4.} Is preprocessing faster on device? & C4,G2 \\
\specialrule{1.5pt}{0pt}{0pt}

\textbf{Aggregation} & \textbf{Agg1.} Are low level events still privacy critical? & G1 \\
\cline{2-3} % Separating line for the question
Low Level $\rightarrow$& \textbf{Agg2.} Are low level events still high-volume? & C1 \\
\cline{2-3}
High Level & \textbf{Agg3.} Can events be build from local context? & C3 \\
\cline{2-3} % Separating line for the question
 Events & \textbf{Agg4.} 
Can sensor/network outages be tolerated?
& C4,C5\\
\specialrule{1.5pt}{0pt}{0pt}

\textbf{Correlation} & \textbf{Cor1.} Does a global notion of case/object ids exist? & C6\\
\cline{2-3} % Separating line for the question
High Level  & \textbf{Cor2.} Is the time synchronized between the nodes? & C5\\
\cline{2-3} % Separating line for the question
Events $\rightarrow$ \newline Event Log & \textbf{Cor3.} Do out of order events violate real-time objectives? & C5,G2 \\
\specialrule{1.5pt}{0pt}{0pt} 
\textbf{Discovery} & \textbf{Dis1.} Is the process model privacy-critical? & C6,G1\\
\arrayrulecolor{black}\cline{2-3} % Separating line for the question
Event Log $\rightarrow$ & \textbf{Dis2.} Does the discovery algorithm benefit from locality? & G2,G3 \\
\arrayrulecolor{black}\cline{2-3} % Separating line for the question
Process Model & \textbf{Dis3.} Does the process mining algorithm require consistent and complete event logs? & C5\\
\specialrule{1.5pt}{0pt}{0pt}
\textbf{Insights} & \textbf{Ins1.} Does insight extraction need advanced hardware? & C4 \\ 
\cline{2-3} % Separating line for the question
Process Model $\rightarrow$ KPIs & \textbf{Ins2.} Can insight extraction tolerate partial results? & C5,G1 \\
\hline
\end{tabular}
\end{table}

\textbf{Application to use-case:}
We demonstrate the \textit{ContinuumConductor} by applying it to the InteGreatDrones project. Thereby, we discuss the strategic placement of computational tasks across the edge-cloud continuum.
In the project context, data is processed across four computing layers: 1) directly at or within sensors (e.g., smart cameras), 2) at edge devices such as mini-computers or gateways, 3) within edge clusters formed by GPU-equipped servers, and 4) in a cloud instance primarily responsible for application and visualization services. Data sources include inertial, distance, and position (e.g., GPS) sensors, as well as cameras, the latter generating the largest data volumes and requiring the most intensive preprocessing.
Specialized tasks such as license plate detection can be executed directly on smart cameras, whereas video streams from cameras mounted on terminal vehicles must be transferred to edge servers for \textit{preprocessing}, as local processing is not feasible (Pre1, Pre4). Since these video streams may contain personal information, anonymization is performed at the edge (Pre2). Given the large volume of video data and the risk of intermittent connectivity in moving vehicles, intelligent filtering of relevant images is also necessary (Pre3). Hence, the decentralized preprocessing in mandatory.
During the \textit{aggregation} step, data such as the position of the container handler and the camera-based identification of containers must be combined. Certain events can be extracted locally, such as a trailer entering the terminal. However, to detect more complex events fusion of proximate sensors is required (Agg3). Single sensor outages may be tolerated (Agg4), hence a distributed protocol for sensor fusion is possible. The low level-events are not necessarily privacy-critical (Agg1), but still occur in high-volume for high frame-rate video streams (Agg2).
In the \textit{correlation} phase, a global case/object ID is ensured, as both trailers and containers possess unique identifiers, which must be recorded at every relevant process step (Cor1). Although timestamps are generally synchronized across devices, the abundance of heterogeneous equipment and potential delays in data synchronization can lead to temporal inconsistencies, highlighting the importance of robust time synchronization mechanisms (Cor2). Handling out-of-order events is critical: Each new entry of a cargo unit into the terminal should be recognized as a distinct case, making the detection and prevention of such events essential (Cor3).
In the \textit{discovery} step, no privacy-critical data remains, as all sensitive information is removed beforehand (Dis1). Since process discovery benefits from consistent and complete event logs (Dis3), and there is no clear advantage to local processing in this case (Dis2), central execution is appropriate.
\textit{Insight extraction} requires a set of complex reasoning algorithms capable of handling conflicting data and determining the most likely outcomes. To achieve this, a comprehensive view of all data sources is necessary (Ins1, Ins2), making centralized processing preferable.

In conclusion \textit{ContinuumConductor} suggests distributed preprocessing near the sensors. For abstraction and correlation distributed processing remains possible for resource efficient computation benefiting from data proximity. For discovery and insight extraction still a central approach is recommended to get an overview on the whole process. 

%The application of the decision framework to this complex real-world project demonstrates that, for every event stream or data source contributing to the final event log, numerous contextual variables—such as data volume, computational hardware, and network infrastructure—must be taken into account. In addition, characteristics like real-time requirements, privacy, and security need to be considered when determining the appropriate placement of the various steps of the process mining pipeline across the cloud–edge continuum.
\section{Conclusion}

This work investigates the challenges and objectives of applying process mining to distributed, high-volume data streams, a field motivated by the InteGreatDrones project. We explore techniques within an edge-cloud continuum and a dedicated process mining pipeline to ensure privacy preservation, real-time insights delivery, and computational efficiency. From this, the \textit{ContinuumConductor} framework has been derived, providing a decision model for centralizing or distributing steps within the process mining pipeline. This framework serves as a foundational discussion point for decentralized process mining within computing infrastructures, necessitating further research into algorithms that balance these identified requirements. Moreover, the question catalog of \textit{ContinuumConductor} needs to be refined by applying it to further scenarios, thereby enhancing its generality and completeness. 

\smallskip
\noindent
\textbf{Acknowledgments}. This work started at Schloss Dagstuhl (Leibniz-Zentrum für Informatik), seminar 25103 ``Process Mining on Distributed Event Sources'' and received funding from the Deutsche Forschungsgemeinschaft (DFG), grant 496119880 and from the German Federal Ministry for Digital and Transport (BMDV) in the funding program Innovative Hafentechnologien II (IHATEC II).

% \newpage

%\begin{credits}
%\subsubsection{\ackname} 
%This work received funding from the Deutsche
%Forschungsgemeinschaft (DFG), grant 496119880.
%The financial support by the Austrian Federal Ministry of Economy, Energy and Tourism, the National Foundation for Research, Technology, and Development and the Christian Doppler Research Association is gratefully acknowledged.
%This work started at Schloss Dagstuhl (Leibniz-Zentrum für Informatik), seminar 25103 “Process Mining on Distributed Event Sources”. 
%\end{credits}

\bibliographystyle{splncs04}
\bibliography{mybibliography}

\end{document}